\documentstyle[letters,referee,amssymb]{mn}

\newif\ifAMStwofonts

\ifoldfss
  \ifCUPmtlplainloaded \else
    \NewTextAlphabet{textbfit} {cmbxti10} {}
    \NewTextAlphabet{textbfss} {cmssbx10} {}
    \NewMathAlphabet{mathbfit} {cmbxti10} {} 
    \NewMathAlphabet{mathbfss} {cmssbx10} {} 
  \fi
  \ifAMStwofonts
    \ifCUPmtlplainloaded \else
      \NewSymbolFont{upmath} {eurm10}
      \NewSymbolFont{AMSa} {msam10}
      \NewMathSymbol{\upi}     {0}{upmath}{19}
      \NewMathSymbol{\umu}     {0}{upmath}{16}
      \NewMathSymbol{\upartial}{0}{upmath}{40}
      \NewMathSymbol{\leqslant}{3}{AMSa}{36}
      \NewMathSymbol{\geqslant}{3}{AMSa}{3E}

       \let\le=\leqslant
       \let\ge=\geqslant
    \fi
  \fi
\fi 

\ifnfssone
  \newmathalphabet{\mathit}
  \addtoversion{normal}{\mathit}{cmr}{m}{it}
  \addtoversion{bold}{\mathit}{cmr}{bx}{it}
  \newmathalphabet{\mathbfit} 
  \addtoversion{normal}{\mathbfit}{cmr}{bx}{it}
  \addtoversion{bold}{\mathbfit}{cmr}{bx}{it}
  \newmathalphabet{\mathbfss} 
  \addtoversion{normal}{\mathbfss}{cmss}{bx}{n}
  \addtoversion{bold}{\mathbfss}{cmss}{bx}{n}
  \ifAMStwofonts
    \ifCUPmtlplainloaded \else
      \UseAMStwoboldmath
      \makeatletter
      \new@mathgroup\upmath@group
      \define@mathgroup\mv@normal\upmath@group{eur}{m}{n}
      \define@mathgroup\mv@bold\upmath@group{eur}{b}{n}
      \edef\UPM{\hexnumber\upmath@group}
      \new@mathgroup\amsa@group
      \define@mathgroup\mv@normal\amsa@group{msa}{m}{n}
      \define@mathgroup\mv@bold\amsa@group{msa}{m}{n}
      \edef\AMSa{\hexnumber\amsa@group}
      \makeatother
      \mathchardef\upi="0\UPM19
      \mathchardef\umu="0\UPM16
      \mathchardef\upartial="0\UPM40
      \mathchardef\leqslant="3\AMSa36
      \mathchardef\geqslant="3\AMSa3E

       \let\le=\leqslant
       \let\ge=\geqslant
    \fi
  \fi
\fi 

\ifnfsstwo
  \DeclareMathAlphabet{\mathbfit}{OT1}{cmr}{bx}{it}
  \SetMathAlphabet\mathbfit{bold}{OT1}{cmr}{bx}{it}
  \DeclareMathAlphabet{\mathbfss}{OT1}{cmss}{bx}{n}
  \SetMathAlphabet\mathbfss{bold}{OT1}{cmss}{bx}{n}
  \ifAMStwofonts
    \ifCUPmtlplainloaded \else
      \DeclareSymbolFont{UPM}{U}{eur}{m}{n}
      \SetSymbolFont{UPM}{bold}{U}{eur}{b}{n}
      \DeclareSymbolFont{AMSa}{U}{msa}{m}{n}
      \DeclareMathSymbol{\upi}{0}{UPM}{"19}
      \DeclareMathSymbol{\umu}{0}{UPM}{"16}
      \DeclareMathSymbol{\upartial}{0}{UPM}{"40}
      \DeclareMathSymbol{\leqslant}{3}{AMSa}{"36}
      \DeclareMathSymbol{\geqslant}{3}{AMSa}{"3E}

       \let\le=\leqslant
       \let\ge=\geqslant
    \fi
  \fi
\fi 

\ifCUPmtlplainloaded \else
  \ifAMStwofonts \else 
    \def\upi{\pi}
    \def\umu{\mu}
    \def\upartial{\partial}
  \fi
\fi

\title[Cosmic Ray Generation by Quasar Remnants]
{Cosmic Ray Generation by Quasar Remnants: Constraints and Implications}
\author[E. Boldt and M. Loewenstein]
{Elihu Boldt$^1$\thanks{E-mail: boldt@lheavx.gsfc.nasa.gov}
and Michael Loewenstein$^{1,2}$\thanks{E-mail: loew@larmes.gsfc.nasa.gov}\\
$^1$Laboratory for High Energy Astrophysics, NASA Goddard Space Flight Center,
Greenbelt, MD, USA\\
$^2$University of Maryland Department of Astronomy, College Park, MD, USA}
\date{Accepted ???. Received ???; in original form ???}

\pagerange{L00-L00}
\pubyear{2000}
\volume{000}

\begin{document}

\maketitle

\label{firstpage}
  
\begin{abstract}
The quasar remnant cores of nearby giant elliptical galaxies NGC 4486
(M87), NGC 1399, NGC 4649 and NGC 4472 are the sites of supermassive
($>10^9$ M$_{\sun}$) black holes. These objects are investigated as to
the viability of the conjecture that they could harbor compact dynamos
capable of generating the highest energy cosmic rays.  For an
accretion process involving an equipartition magnetic field near the
event horizons of the underlying putative spun-up black holes, the
energy achievable in accelerating protons could well be $\ge 10^{20}$
eV for all these when only considering the drag induced by curvature
radiation.  Estimates of the SED (spectral energy distribution) of
ambient core photons lead to the conclusion that the energy losses
arising from photo-pion production in proton collisions with these
target photons are relatively small for all but M87. For M87, the
ambient photon field is likely to be a limiting factor.  Accretion
rates of $\sim 1\ {\rm M}_{\sun}\ {\rm yr}^{-1}$, comparable to the
Bondi rates and to the stellar mass loss rates, are associated with
($> 10^{20}$ eV) cosmic ray generation in the other
(electromagnetically dark) galactic core sites.  If these sites are
found to be sources of such cosmic rays, it would suggest the presence
of a global inflow of interstellar gas all the way into the center of
the host galaxy.
\end{abstract}

\begin{keywords}
acceleration of particles -- black hole physics -- cosmic rays -- 
galaxies: nuclei -- accretion.
\end{keywords}

\section{Introduction}

The massive dark objects (MDOs) at the centers of present-epoch giant
elliptical galaxies appear to be the supermassive black hole remnants
of earlier quasar activity (McLure et al. 1999; McLeod, Rieke \&
Storrie-Lombardi 1999; Salucci et al. 1999; Richstone et al. 1998;
Boldt \& Leiter 1995; Chokshi \& Turner 1992). It has been proposed
that the highest energy cosmic ray particles, those above the `GZK'
limit (Greisen 1966; Zatsepen \& Kuz'min 1966), could be accelerated
by the effective electromotive force (emf) generated near the event
horizon of spinning supermassive ($\gtrsim 10^9{\rm M}_{\sun}$) black
holes at the apparently dormant quasar remnant cores of nearby massive
elliptical galaxies \cite{bg}. While an emf up to $10^{21}$ volts or
somewhat greater would appear to be possible, the drag arising from
curvature radiation \cite{lev} limits the energy to be attained by a
charged particle accelerated by this field; for the four cases
considered here (i.e., NGC 4486, NGC 1399, NGC 4649 \& NGC 4472) the
suppression factor for a proton is estimated to be $\sim 6 -
9$. Acceleration to energies $\ge 10^{20}$ eV would then still be
feasible. However, during the final phase of the acceleration process
energy losses due to inelastic collisions with ambient photons, those
dominated by photo-pion production, must be considered.  Therefore, we
address in some detail the major concern that recent radio data on
low-luminosity AGNs could imply a photon number density at the nuclear
core that is so high as to preclude ultra-relativistic processes of
particle acceleration and escape, particularly as they pertain to the
immediate vicinity of the putative supermassive black hole \cite{b00}.

Since $\gamma=10^{11}$ for a $10^{20}$ eV proton, ambient photons at
$h\nu > 15\times 10^{-4}$ eV (i.e., \@$\nu>360$ GHz; $\lambda<0.83$
mm) would appear in the nucleon's proper frame as $\gamma$-rays $>150$
MeV, above the critical threshold for energy losses via inelastic
collisions involving pion production (cf., Stecker 1968; Hill \&
Schramm 1985; Stecker \& Salamon 1999).  Using recent estimates of the
SED (spectral energy distribution) at the core of M87 (NGC 4486), the
central (and dominant) giant elliptical galaxy in the Virgo Cluster
(Reynolds et al. 1996; Ho 1999; Di Matteo et al.  2000), we find that
the corresponding radiation length of such energetic protons might
well be significantly smaller than the Schwarzschild radius, thereby
precluding the acceleration to such a high energy.  Hence, if M87 is
indeed found to be a source of the highest energy cosmic rays, as
commonly conjectured (Biermann \& Strittmatter 1987; Biermann 1999;
Farrar \& Piran 2000), the underlying accelerator would then have to
be other than the compact dynamo considered here. However, for other
nearby giant elliptical galaxies harboring apparently dormant
supermassive ($>10^9{\rm M}_{\sun}$) black holes, in particular for
the remaining three such objects addressed by Di Matteo et al. (1999,
2000), the radiation lengths are estimated to be larger than their
Schwarzschild radii. These include NGC 1399, the central giant
elliptical in the Fornax Cluster and two other giant elliptical
galaxies in Virgo (i.e., NGC 4649 and NGC 4472). The dynamo
characteristics expected for the associated compact cores (i.e.,
$B$-field, achievable energy) are derived from independent estimates
of the black hole mass, Bondi accretion rate and SED environment. It
has been pointed out that TeV $\gamma$-ray curvature radiation is a
necessary consequence of cosmic ray generation by such black hole
dynamos \cite{lev}; these giant elliptical galaxies would be
worthwhile targets for the observation of such electromagnetic
radiative signature.

\section{Dynamo Characteristics}

The dynamo's emf (V) is generated by the black hole induced rotation
of externally supplied magnetic field lines threading the horizon
\cite{bz}. If $B$ is the ordered poloidal field near the hole, $V\sim
aB$, where $a$ is the hole's specific angular momentum; for a hole
mass $M$, $a\le M$ (e.g., $a=M$ for an extreme Kerr hole). In
astrophysical units \cite{z78}:
\begin{equation}
V=9\times 10^{20}(a/M)B_4M_9\ {\rm volts},
\end{equation}
where $B_4\equiv B/(10^4 {\rm G})$ and $M_9\equiv M/(10^9{\rm
M}_{\sun})$.

The energy density of the magnetic field near the event horizon is
expected to be in equipartition with the rest mass energy density of
accreting matter \cite{k99}. In terms of an Advection Dominated
Accretion Flow (ADAF) model (cf., Di Matteo et al. 1999) this is to be
identified with the regime where the gas pressure is half the total
(i.e, $\beta=1/2$). Under this assumption,
\begin{equation}
B_4=1.33{M_9}^{-1}{\dot{\rm M}}^{1/2},
\end{equation}
where $\dot{\rm M}$ is the accretion rate $dM/dt$ in ${\rm M}_{\sun}\ {\rm
yr}^{-1}$.

The magnetic field ($B$) to be associated with the nuclei of those
elliptical galaxies considered here is obtained via equation 2 using
the Bondi accretion rates $\dot{\rm M}_{\rm Bondi}$ estimated by Di
Matteo et al. \shortcite{d00} and the black hole masses determined by
Magorrian et al. \shortcite{m98}.

Noting that the hole's radius is fixed by its mass, the emf given by
equation 1 may be expressed directly in terms of the accretion
rate. The maximum emf ($V$), that corresponding to $a/M$ close to
unity, is then given by:
\begin{equation}
V=1.2\times 10^{21}{\dot{\rm M}}^{1/2}\ {\rm volts}.
\end{equation}
Except where noted, we take $\dot{\rm M}=\dot{\rm M}_{\rm Bondi}$.

The energy ($E$) to be attained in this electric field is limited by
the drag arising from curvature radiation induced by the magnetic
field \cite{lev}.  From equation 5 in Levinson (op. cit.), we obtain
that, for a proton (charge $e^+$), the suppression ratio is given by:
\begin{equation}
E/[e(V)]\approx [(50M_9)^{-1/2}{B_4}^{-3/4}]r^{1/2},
\end{equation}
where $r$ is the magnetic field curvature in units of the
Schwarzschild radius.  For $r\approx 1$ and ${\dot{\rm M}}\approx
(0.1-10)\ {\rm M}_{\sun}\ {\rm yr}^{-1}$, we note (from equations
2--4) that $E=(1.0-1.8)\times 10^{20}{M_9}^{1/4}$ eV. Although the
energy possible for a heavier nucleus could be substantially greater,
such a particle (e.g., Fe) launched at $\ge 10^{21}$ eV would be
disrupted into its constituent nucleons after traveling only 20 Mpc
(e.g, Cronin 1997). By contrast, a proton starting with $10^{21}$ eV
would maintain about one-third of its initial energy after traversing
the same distance.

During the final phase of the acceleration process energy loss due to
photo-pion production in collisions with ambient photons becomes a
relatively important effect . The associated radiation length
($\Lambda$) relative to $R$ (the radius of core emission) is given by
\begin{eqnarray}
\Lambda/R & = & c\pi R/(\langle K\sigma\rangle Q) \\
          & = & (278/\langle K\sigma '\rangle)(R/R_S)M_9(Q/10^{53}\ 
{\rm s}^{-1})^{-1}, \nonumber
\end{eqnarray}
where $\sigma$ is the proton photo-pion production cross-section,
$\sigma '$ its value in microbarns ($10^{-30}$ cm$^2$), $R_S$ is the
Schwarzschild radius, $K\equiv \langle E({\rm loss})\rangle/E({\rm
initial})$ is the inelasticity in a single collision \cite{s68} and
$Q$ is the core emission rate (photons s$^{-1}$) for electromagnetic
radiation at $\nu>360$ GHz, given by
\begin{equation}
Q=h^{-1}\int\nu^{-1}L_{\nu}d\nu
\end{equation}
where $h$ is the Planck constant and $L_{\nu}=4\pi D^2F_{\nu}$ for a
source of spectral density $F_{\nu}$ at distance $D$. We note
\cite{c98} that, for the regime of interest here,
\begin{equation}
\langle K\sigma '\rangle\equiv 
\left[\int(K\sigma '(dQ/d\nu)d\nu\right]/Q<120\ {\rm microbarns}.
\end{equation}

\begin{table*}
\begin{minipage}{140mm}
\caption{Black Hole Galactic Nuclei: Candidate Cosmic Ray Sources}
\begin{tabular}{@{}lccccc}
 Host Galaxy & & NGC 4486 & NGC 1399 & NGC 4649 & NGC 4472 \\
 Host Cluster & & Virgo & Fornax & Virgo & Virgo \\
 & & & & & \\
 Distance~($D$) & (Mpc) & 18 & 29 & 18 & 18 \\
 & & & & & \\
 Black Hole Mass~($M$) & ($10^9$ M$_{\sun}$) & 3.6 & 5.2 & 3.9 & 2.6 \\
 & & & & & \\
 $\dot{\rm M}_{\rm Bondi}$ & (${\rm M}_{\sun}\ {\rm yr}^{-1}$) & 1.5 & 3 &
 1.4 & 0.7 \\
 & & & & & \\
 Stellar Bulge Mass~(M$_{\rm Bulge}$) & ($10^{12}$ M$_{\sun}$) & 0.82 & 0.32 
 & 0.54 & 0.84 \\
 & & & & & \\
 $L_{\rm Obs}/L_{\rm Bondi}$ & & $10^{-5}$ & $2\times 10^{-6}$ & 
 $3\times 10^{-5}$ & $10^{-5}$ \\
 & & & & & \\
 Magnetic Field~($B$) & 
($10^4$ G) & 0.45 & 0.44 & 0.40 & 0.43 \\
 & & & & & \\
 emf~($V$) & ($10^{20}$ volts) & 15 & 21 & 14 & 10 \\
 & & & & & \\
 $r^{-1/2}E$ & ($10^{20}$ eV) & -- & 2.4 & 2.0 & 1.7 \\
 & & & & & \\
 Obs. Radio Freq.~($\nu_{\rm obs}$) & (GHz) & 100 & 43 & 43 & 43 \\
 & & & & & \\
 $\nu L_{\nu}$($\nu_{\rm obs}$) & ($10^{38}$ erg s$^{-1}$) &
 32 & 6.7 & 2.0 & 3.3 \\
 & & & & & \\
 $\langle \nu L_{\nu}$(360 GHz)$\rangle$ & ($10^{38}$ erg s$^{-1}$) &
 47 & 13 & 3.8 & 6.2 \\
 & & & & & \\
 $\nu L_{\nu}$($7\times 10^8$ GHz) & ($10^{38}$ erg s$^{-1}$) &
 -- & $<13$ & -- & -- \\
 & & & & & \\
 $\langle Q(\lambda<0.83 {\rm mm})\rangle$ & ($10^{53}$ photon s$^{-1}$) &
 $\sim 28$ & $<7.7$ & $<2.3$ & $<3.7$ \\
 & & & & & \\
 $(R_S/R)\Lambda/R$ & & $\sim 0.3$ & $>2$ & $>4$ & $>2$ \\
 \end{tabular}

\medskip
The black hole and galactic bulge masses are from Magorrian et
al. \shortcite{m98}.  The distances, Bondi accretion rates and $L_{\rm
Obs}/L_{\rm Bondi}$ ratios are taken from Di Matteo et
al. \shortcite{d00}.  Radio observations are at 0.3 cm for NGC
4486/M87 (Reynolds et al.  1996; B{\"a}{\"a}th et al. 1992) and 0.7 cm
for the three other sources (Di Matteo et al. 1999, 2000). X-ray
observations with the {\it Chandra} Observatory at $h\nu=3$ keV
\cite{loe} correspond to $\nu=7\times 10^8$ GHz.
\end{minipage}
\end{table*}

\section{Constraints}

The key characteristics of the four dynamo candidates investigated
here are summarized in Table 1. The tabulated values for $\nu L_{\nu}$
at $\nu=360$ GHz were obtained from $F_{\nu}$ measurements at lower
frequencies under the assumption that $F_{\nu}\propto \nu^{-0.7}$,
consistent with the radio spectra observed in the vicinity of 43 GHz
(Di Matteo et al. 1999, 2000; Reynolds et al. 1996).  The core
emission rate ($Q$ photons s$^{-1}$) at $\lambda<0.83$ mm ($\nu>360$
GHz) from the accretion flow associated with each of the four
supermassive black holes is estimated via equation 6 under the
assumption that $F_{\nu}\propto\nu^{-0.7}$ at higher frequencies as
well. For M87 this is based on a high resolution ($10^{-4}$ arcsec)
VLBI measurement at 100 GHz (Reynolds et al. 1996; B{\"a}{\"a}th et
al. 1992). For the other three sources we rely on VLA measurements at
43 GHz (Di Matteo et al. 1999, 2000). Because these latter
measurements allow more of a contribution from the underlying galaxy
and weak jets, and two (NGC 4472 \& NGC 4649) do show some extended
emission, they are taken to be upper limits to the flux from the
compact cores of these galaxies.  Comparisons of these extrapolated
spectra with the observed limits shown in Figure 2 of Di Matteo et
al. \shortcite{d00} are consistent with these estimates being upper
limits to $L_{\nu}$ at $\nu\ge 360$ GHz, although less clear for M87.
The presence of inner jets associated with any of these potential
dynamos would imply vacuum breakdown if created by the mechanism
suggested by Blandford \& Znajek (1977). In this situation, the
feasibility of our compact cosmic ray generator demands that such a
jet ejection process be episodic.

The magnetic field ($B$) is obtained from the Bondi accretion rate and
black hole mass via equation 2. The emf is calculated from the Bondi
accretion rate via equation 3. Considering the drag arising from
curvature radiation \cite{lev} the energy ($E$) to be attained by a
proton accelerated by this emf is estimated via equation 4. During the
final phase of the acceleration process energy losses due to
collisions with ambient photons are dominated by photo-pion
production. The associated radiation length ($\Lambda$) relative to
$R$ listed in Table 1 is obtained from equation 5, where $R$ is the
radius of core emission and $R_S$ is the Schwarzschild radius.

Lack of microwave data for these sources prevents any direct
confirmation that our power-law extrapolation from the radio
(centimeter) band does in fact provide an upper limit to their
emission in the entire relevant submillimeter band at $\lambda<0.83$
mm. As a result, we depend on the ADAF models prescribed by Di Matteo
et al. \shortcite{d00} to provide us with the template needed for
evaluating our power-law extrapolation. For M87 our extrapolation
equals or exceeds $L_{\nu}$ expected for all their models except the
`no-wind' ADAF; see Figure 2a in their paper. For the three other
sources our power-law definitely exceeds what is expected for all
models except the `no-wind' ADAF; see Figures 2b, 2d \& 2e. Radio data
exhibited by Di Matteo et al. \shortcite{d00} at longer wavelengths,
at least for NGC 4649 and NGC 4472, appear to rule out the no-wind
ADAF model in favor of models with winds that have less emission at
frequencies $>360$ GHz than those based on the simple extrapolations
presented in Table 1. Moreover, analysis of {\it Chandra} X-ray
Observatory data for NGC 1399 (see below, and Loewenstein et al. 2000)
places an upper limit on $\nu L_{\nu}$ at 3 keV ($\sim 7\times
10^{17}$ Hz) of $1.3\times 10^{39}$ erg s$^{-1}$ (where, for
consistency, we adopt the distance to NGC 1399 of 29 Mpc used by Di
Matteo et al. 2000) with comparable limits throughout the 0.3--3.0 keV
X-ray energy range. A preliminary estimate of the detected UV flux
from a nuclear point source yields a comparable value of $\nu L_{\nu}$
at $\nu\sim 2\times 10^{15}$ Hz \cite{o00}. Simply extrapolating the
highest energy radio point (at 43 GHz) through the X-ray upper limit
reduces the value of the photon emission rate $Q$ in Table 1 by about
a factor of 3 for NGC 1399, thereby increasing the associated lower
limit to $\Lambda/R$ threefold.

\section{Implications} 

The vicinities of spun-up supermassive black holes in dormant
elliptical galaxy nuclei can provide an acceleration mechanism
sufficient to explain observations of the highest energy cosmic rays,
and meet the further criterion of being numerous in the nearby ($<50$
Mpc) universe defined by the GZK cutoff for the highest energy
particles. Since few, if any, other astrophysical sites in the
present-epoch are as feasible, the comparison of the following
predictions with cosmic ray data of improved spatial and spectral
statistics could determine whether models invoking new particles or
topological defects need be considered more carefully.  Conversely,
confirmation of the present model has far-reaching astrophysical
implications; e.g, it would indicate continued spin-up and growth of
supermassive black holes via accretion.

The angular distribution of ultra-high energy (UHE) cosmic ray sources
provides the cleanest test of a supermassive black hole origin, as the
most massive black holes lie in the nuclei of elliptical
galaxies. Scattering by intracluster fields (microgauss $B$) may
destroy this association for galaxies in clusters; however, TeV
$\gamma$-rays created via curvature radiation \cite{lev} would
preserve the correlation. Also, one would not expect otherwise active
galaxies, with their higher central photon densities, to produce these
cosmic rays. Finally, under the equipartition assumption the emf
should be proportional to the square root of the rate of accretion
into the nucleus, which should be well approximated by the Bondi rate
(Brighenti \& Mathews 1999; Quataert \& Narayan, 2000), i.e.
$V\propto \rho_{\rm gas}^{1/2}T_{\rm gas}^{-3/4}M$, where $\rho_{\rm
gas}$ and $T_{\rm gas}$ at the accretion radius are measurable using
the {\it Chandra} X-ray Observatory.

The estimates of the radiation lengths in Table 1 rest on the
extrapolation of the radio spectrum through 360 GHz assuming a
$\nu^{-0.7}$ energy spectrum. A significantly flatter (or inverted)
slope would imply a photon number density that is prohibitively high
for particle acceleration to $>10^{20}$ eV; and, indeed this seems to
be the case for many low-luminosity AGN \cite{h99}. However, with the
exception of M87, the galaxies considered here may not be members of
the same population, as they lack detectable optical emission lines
\cite{hfs}.

The {\it Chandra} limit on the 3 keV nuclear emission for NGC 1399 is
derived using the extracted spectrum in a 3.5 arcsec diameter
aperture.  Two-component models including thermal (hot ISM) and
power-law components are fit to the data and 90\% confidence upper
limits to the power-law flux derived. This upper limit is robust as it
accounts for all the observed emission above 2 keV, and is
conservative since image analysis strongly suggests an even lower
upper limit to the flux originating in any nuclear point source
\cite{loe}. The hard component attributed by Di Matteo et
al. \shortcite{d00} to an ADAF associated with the nuclear
supermassive black hole has now been resolved with {\it Chandra} into
an extended distribution of discrete sources \cite{a00}. The VLA radio
observations of NGC 4649, NGC 4472, and NGC 1399 correspond to a
similar angular resolution as that of {\it Chandra} and should be
regarded as upper limits as well since there may be contributions from
inner-jet emission [the spectra in Ho \shortcite{h99} are derived from
higher resolution VLBI observations].  Thus these galaxies are
quiescent even relative to LINERS and other low luminosity AGN, and it
is plausible that cosmic ray energy losses by ambient photons are
unimportant for these three galaxies. We note that M87, a likely
exception, displays more powerful activity at all energies.

If the {\it Chandra} results for NGC 1399 are generalizable, the
nuclear X-ray emission from elliptical galaxies is well below that
predicted by models of ADAFs accreting at the Bondi rate.  However,
for the specific cases considered in this paper the Bondi accretion
rate that provides sufficient compression of the magnetic field to
generate an emf corresponding to the most energetic cosmic rays is of
the same order as the integrated stellar mass loss rate.  $\dot{\rm
M}_{\rm Bondi}/{\rm M}_{\rm Bulge}=$ 1.8, 9.4, 2.6 and 0.83 $\times
10^{-12}\ {\rm yr}^{-1}$ for M87, NGC 1399, NGC 4649 and NGC 4472,
respectively (Table 1).  These are comparable to the specific mass
return rate calculated for the stellar population in elliptical
galaxies (e.g., Mathews 1989), suggesting the presence of a global
inflow of interstellar gas that persists from large galactic radii all
the way into the very center of the galaxy. Thus while (hot)
protogalactic gas may be the source for the initial rapid growth and
quasar-epoch fueling of massive black holes \cite{nf}, the stellar
population of the galaxy at large may be responsible for `feeding the
monster' \cite{g79} during the present era: an era where the quieter
power output may very well be characterized by high energy particles
and their associated gamma radiation.

\section{Outlook}

Comprehensive investigations of nearby giant elliptical (\& S0)
galaxies indicate that most harbor a massive dark object (MDO) at
their centers \cite{m98}. Studies of AGN evolution (Chokshi \& Turner
1992; Fabian \& Iwasawa 1999; Salucci et al. 1999) and the X-ray
background of accretion-powered radiation \cite{bl} conclude that the
largest of these MDOs are associated with supermassive black holes
which are the apparently dormant remnants of previously active
quasars. It has been proposed that these accretion-fed spinning dark
objects are latent dynamos, sufficient for producing the highest
energy cosmic rays \cite{bg}. The accretion rate for such a dynamo is
no greater than the mass loss rate estimated for giant elliptical
galaxies \cite{m89}.  The scatter and uncertainty in the ratios of
black hole to bulge mass and of bulge mass to optical luminosity, and
the inhomogeneity of the galaxy distribution within 50 Mpc, make it
virtually impossible to precisely quantify the expected underlying
number of UHE cosmic ray sources. However, we note that the number of
galaxies within 50 Mpc that have bulges sufficiently luminous ($L_{\rm
Bulge}>10^{10} L_{B\sun}$) to be potential sites of $>3\times 10^8\
{\rm M}_{\sun}$ black holes is on the order of $10^3$ (Magorrian 1999,
Marinoni et al. 1999).  Assuming an IGM $B$ field of coherence length
$\sim 1$ Mpc and magnitude $\sim 1$ nanogauss, a UHE proton $>10^{20}$
eV originating within 50 Mpc would have an arrival direction aligned
sufficiently well with its origin for avoiding confusion among
candidate sources \cite{mgh}, although it has been suggested that the
relevant B field could be larger than a nanogauss \cite{fp}. For a UHE
proton $<10^{21}$ eV originating from a source deep within a rich
cluster of galaxies the Larmor radius (for the associated microgauss B
field) is much less than the cluster radius \cite{ss8}; hence the
emerging protons would appear to be coming from an extended source the
size of the cluster.  Fortunately, most suitably massive giant
elliptical galaxies reside outside of clusters (Magorrian et al. 1998;
Burstein 1999). If the `local' extragalactic magnetic field is much
less than a microgauss, then an angular resolution of about a degree
should be good enough for the OWL/Airwatch air-shower observatory
\cite{s98} to establish a correlation between candidate ellipticals
and UHE cosmic ray events, at least two-thirds of the time (the
remainder would be cluster associated). Those in clusters would then
be located via their TeV $\gamma$-ray curvature radiation. In
particular, of the six such giant ellipticals addressed by Di Matteo
et al. \shortcite{d00} we find that, after considering principal
constraints, at least half could well be harboring supermassive black
hole dynamos capable of producing cosmic rays more energetic than
$10^{20}$ eV and be identifiable via their TeV $\gamma$-radiation,
e.g., with the Whipple Observatory at an angular resolution sufficient
for isolating specific galaxy emission \cite{w96}.

\section*{Acknowledgments}

We thank R. D. Blandford, P. Biermann, D. Burstein, J. Finley,
G. Giuricin, W. Landsman, A.  Levinson, G. Medina-Tanco, J. Magorrian,
R. Mushotzky, F. W. Stecker and R.  Streitmatter for informative
discussions and are especially grateful to O.  W. Greenberg,
J. Linsley and J. F. Ormes for their considerable encouragement.

\bsp 

\label{lastpage}

\end{document}